\documentstyle[12pt]{article}
\newcommand{\bea}{\begin{eqnarray}}
\newcommand{\eea}{\end{eqnarray}}
\newcommand{\beq}{\begin{equation}}
\newcommand{\eeq}{\end{equation}}

\def\mn{\medbreak\noindent}
\def\msbar{\ifmmode{\overline{\rm MS}} \else{$\overline{\rm MS}$} \fi}
\def\drbar{\ifmmode{\overline{\rm DR}} \else{$\overline{\rm DR}$} \fi}
\def\st{\ifmmode{\tilde{t}} \else{$\tilde{t}$} \fi}
\def\sb{\ifmmode{\tilde{b}} \else{$\tilde{b}$} \fi}
\def\sq{\ifmmode{\tilde{q}} \else{$\tilde{q}$} \fi}
\def\sg{\ifmmode{\tilde{g}} \else{$\tilde{g}$} \fi}
\def\bbar{\ifmmode{\bar{b}} \else{$\bar{b}$} \fi}
\def\tbar{\ifmmode{\bar{t}} \else{$\bar{t}$} \fi}
\def\qbar{\ifmmode{\bar{q}} \else{$\bar{q}$} \fi}
\def\ksla{{k \hspace{-2mm} \slash}}

\newcommand{\gsim}{\;\raisebox{-0.9ex}
           {$\textstyle\stackrel{\textstyle >}{\sim}$}\;}

\begin{document}
\pagestyle{empty}
\vspace*{-3cm}
\begin{flushright}
UWThPh-1995-35\\
HEPHY-PUB 631/95\\
TGU-16\\
ITP-SU-95/04\\
KEK-TH-451\\
hep-ph/9511385\\
\vspace{0.3cm}
November, 1995
\end{flushright}

\vspace{1cm}
\begin{center}
\begin{Large} \bf
QCD corrections to the decay $H^+ \rightarrow t \bar{b}$\\
in the\\
Minimal Supersymmetric Standard Model\\
\end{Large}
\end{center}
\vspace{10mm}

\begin{center}
\large A.~Bartl,$^1$ H.~Eberl,$^2$ K.~Hidaka,$^3$ T.~Kon,$^4$\\
W.~Majerotto$^2$, and Y.~Yamada$^5$\footnote{Present address: Physics
Department, University of Wisconsin, Madison, WI 53706, USA}
\end{center}
\vspace{0mm}

\begin{center}
\begin{tabular}{l}
$^1${\it Institut f\"ur Theoretische Physik, Universit\"at Wien, A-1090
Vienna, Austria}\\
$^2${\it Institut f\"ur Hochenergiephysik der \"Osterreichische Akademie
der Wissenschaften,}\\
{\it A-1050 Vienna, Austria}\\
$^3${\it Department of Physics, Tokyo Gakugei University, Koganei,
Tokyo 184, Japan}\\
$^4${\it Faculty of Engineering, Seikei University, Musashino, Tokyo 180,
Japan}\\
$^5${\it Theory Group, National Laboratory for High Energy Physics (KEK)},\\
{\it Tsukuba, Ibaraki 305, Japan}\\
\end{tabular}
\end{center}
\vspace{10mm}

\begin{abstract}
\begin{small}
\baselineskip=14pt
We present a complete calculation of the ${\cal O}(\alpha_s)$ QCD corrections
to the width of the decay $H^+ \to t \bar{b}$ within the Minimal
Supersymmetric Standard Model. We find that the QCD corrections are quite
important, and that the supersymmetric QCD corrections (due to gluino,
$\st$ and $\sb$ exchange) can be comparable to or even larger than the standard
QCD corrections in a large region of the supersymmetric parameter space. This
is mainly due to the effect of large left--right mixings of stop ($\st$) and
sbottom ($\sb$).
This could significantly affect the phenomenology of the $H^+$ search.
\end{small}
\end{abstract}


\newpage
\pagestyle{plain}
\setcounter{page}{2}
\baselineskip=21pt

\section{Introduction}
The existence of a charged Higgs boson $H^+$ would be a clear indication
that the Standard Model must be extended. For example,
the Minimal Supersymmetric
Standard Model (MSSM) \cite{1} with two Higgs doublets predicts the
existence of five
physical Higgs bosons $h^0, H^0, A^0$, and $H^\pm$ \cite{2,3}. If all
supersymmetric (SUSY) particles are heavy enough, $H^+$ decays dominantly
into $t \bar{b}$ above the $t \bar{b}$ threshold \cite{2,4}. In
refs.~\cite{5,6}
all decay modes of $H^+$ including the SUSY--particle modes were studied in the
case that the SUSY--particles are relatively light: it was found that the
$t \bar{b}$ mode remains important even in this case (though the $\tilde t
\bar{\tilde b}$ mode can be dominant in a wide range of the MSSM parameters).
Thus it is important to calculate the QCD corrections to the $t \bar{b}$
mode as they could significantly affect the phenomenology of the $H^+$~search.
The standard QCD corrections to the $t \bar{b}$ mode
were already calculated \cite{7}:
they can be large (+10\% to --50\%). There also exist calculations of the
SUSY--QCD corrections within the MSSM \cite{8,9}. However, in ref.~\cite{8} the
squark--mixing was neglected. The calculation in ref.~\cite{9} is
incomplete as the wave function and mass renormalizations were omitted. A
calculation of SUSY--QCD corrections to the related process $t \to H^+ b$
was done recently in ref.~\cite{9a}.\\
In this paper we present a complete calculation of the
${\cal O}(\alpha_s)$ QCD corrections to the width of $H^+ \to t \bar{b}$ within
the MSSM. We include the left--right mixings of both the $\st_{L,R}$
squarks and the $\sb_{L,R}$ squarks.
We adopt the on--shell renormalization scheme.

\section{QCD one--loop contributions}

The one--loop corrected  amplitude of the decay $H^+(p) \to t(k_t)
\bbar(k_{\bbar})\,\,(p = k_t + k_{\bbar})$ can be written as
\beq
{\cal M}=i\tbar(Y_1P_R+Y_2P_L)b
\eeq
with $P_{R,L} = \frac 1 2 (1 \pm \gamma_5)$ and the one--loop corrected
couplings:
\beq \label{2}
Y_i = y_i+\delta Y_i^{(g)} + \delta Y_i^{(\sg)} \;\; (i=1,2)\, ,
\eeq
where $y_i$ are the tree--level couplings corresponding to Fig.~1a:
\bea
y_1&=&\frac{g}{\sqrt{2}m_W}m_b\tan\beta=h_b\sin\beta \nonumber\,, \\
y_2&=&\frac{g}{\sqrt{2}m_W}m_t\cot\beta=h_t\cos\beta \, ,
\label{treecouplings}\eea
with $g$ being the SU(2) coupling.
$\delta Y_i^{(g)}$ and $\delta Y_i^{(\sg)}$ are the contributions from
gluon and gluino exchanges, respectively (as shown in Figs.~1b and 1c).\\
The tree--level decay width is given by:
\beq \label{gammatree}
\Gamma^{\mbox{tree}}(H^+\to t\bbar)
=\frac{N_C\kappa}{16\pi m_{H^+}^3}[(m_{H^+}^2-m_t^2-m_b^2)(y_1^2+y_2^2)
-4m_t m_b y_1 y_2 ]\, ,
\eeq
where $\kappa=\kappa(m_{H^+}^2, m_t^2, m_b^2 )$,
$\kappa(x,y,z)\equiv ((x-y-z)^2-4yz)^{1/2}$, and $N_C=3$.\\

The vertex corrections due to gluon and gluino exchanges at the vertex
(Fig.~1b), $\delta Y_i^{(v,g)}$ and $\delta Y_i^{(v,\sg)}$, respectively,
are given by:
\bea \label{vertexcorr}
\delta(Y_1P_R+Y_2P_L)^{(v,g)}&=&\frac{\alpha_sC_F}{4\pi}\left\{
2[B_0(m_t^2, 0, m_t^2)+B_0(m_b^2, 0, m_b^2) -r \right. \nonumber \\
&&-(m_{H^+}^2-m_t^2-m_b^2)C_0(\lambda^2, m_t^2, m_b^2)](y_1P_R+y_2P_L)
\nonumber \\
&& - 2 m_t C_1(\lambda^2, m_t^2, m_b^2)[(m_ty_1+m_by_2)P_R+(m_ty_2+m_by_1)P_L]
\nonumber \\
&& - 2 m_b C_2(\lambda^2, m_t^2, m_b^2) [(m_ty_2+m_by_1)P_R+(m_ty_1+m_by_2)P_L]
\left.\right\}\nonumber\, ,\\
\delta(Y_1P_R+Y_2P_L)^{(v,\sg)}&=&\frac{\alpha_sC_F}{4\pi}\left\{\right.
2G_{ij}\left[-m_{\sg}C_0(m_{\sg}^2, m_{\st_i}^2,  m_{\sb_j}^2)
\{(\alpha_{LR})_{ij}P_R+(\alpha_{RL})_{ij}P_L\}\right. \nonumber\\
&& + m_t C_{1}(m_{\sg}^2, m_{\st_i}^2,  m_{\sb_j}^2)
\{(\alpha_{LL})_{ij}P_L+(\alpha_{RR})_{ij}P_R\}  \nonumber\\
&& \left.\left. + m_b C_{2}(m_{\sg}^2, m_{\st_i}^2,  m_{\sb_j}^2)
\{(\alpha_{LL})_{ij}P_R+(\alpha_{RR})_{ij}P_L\}
\right]\right\}\, .
\eea
with $C_F = 4/3$ and
\[
\alpha_{LL}=\left( \begin{array}{rr}
\cos\theta_\st\cos\theta_\sb & -\cos\theta_\st\sin\theta_\sb \\
-\sin\theta_\st\cos\theta_\sb &
\sin\theta_\st\sin\theta_\sb \end{array} \right) , \;\;\;
\alpha_{LR}=\left( \begin{array}{rr}
-\cos\theta_\st\sin\theta_\sb & -\cos\theta_\st\cos\theta_\sb \\
\sin\theta_\st\sin\theta_\sb &
\sin\theta_\st\cos\theta_\sb \end{array} \right) , \]
\beq
\alpha_{RL}=\left( \begin{array}{rr}
-\sin\theta_\st\cos\theta_\sb & \sin\theta_\st\sin\theta_\sb \\
-\cos\theta_\st\cos\theta_\sb &
\cos\theta_\st\sin\theta_\sb \end{array} \right) , \;\;\;
\alpha_{RR}=\left( \begin{array}{rr}
\sin\theta_\st\sin\theta_\sb & \sin\theta_\st\cos\theta_\sb \\
\cos\theta_\st\sin\theta_\sb &
\cos\theta_\st\cos\theta_\sb \end{array} \right) .
\eeq
$G_{ij}$ are the tree--level couplings of $H^+$ to
$\st_i \bar{\sb}_j\, (i,j = 1,2)$ reading:
\bea \label{7}
G_{ij}&=& \frac{g}{\sqrt{2}m_W}
R^{\st}\left( \begin{array}{cc}
m_b^2\tan\beta+m_t^2\cot\beta-m_W^2\sin 2\beta & m_b(A_b\tan\beta+\mu) \\
m_t(A_t\cot\beta+\mu) &
2m_tm_b/\sin 2\beta \end{array} \right)(R^{\sb})^{\dagger}\, . \nonumber\\
\eea
Here $R^{\sq}$ $(\sq = \st$ or $\sb)$ is the $\tilde q$--mixing matrix
\beq \label{5}
R^{\sq}_{i\alpha}=\left(
\begin{array}{cc}\cos\theta_\sq & \sin\theta_\sq \\
-\sin\theta_\sq & \cos\theta_\sq \end{array}\right)
\quad (i = 1,2; \, \alpha = L, R)
\eeq
relating the squark states $\sq_L$ and
$\sq_R$ to the mass--eigenstates $\sq_1$ and
$\sq_2$ $(m_{\sq_1} < m_{\sq_2})$: $\sq_i=R^{\sq}_{i\alpha}\sq_{\alpha}$.
$R^{\sq}$ diagonalizes the
squark mass matrix \cite{3}:
\beq
\left( \begin{array}{cc}m_{LL}^2 & m_{LR}^2 \\ m_{RL}^2 & m_{RR}^2
\end{array} \right)=
(R^{\sq})^{\dagger}\left( \begin{array}{cc}m_{\sq_1}^2 & 0 \\ 0 &
m_{\sq_2}^2 \end{array}
\right)R^{\sq}, \eeq
where
\bea
m_{LL}^2 &=& M_{\tilde{Q}}^2+m_q^2+m_Z^2\cos 2\beta (I_q-Q_q\sin^2\theta_W), \\
m_{RR}^2 &=& M_{\{\tilde{U},
\tilde{D}\}}^2+m_q^2+m_Z^2\cos 2\beta Q_q\sin^2
\theta_W, \\
m_{LR}^2=m_{RL}^2 &=& \left\{ \begin{array}{ll}
m_t(A_t-\mu\cot\beta) & (\sq=\st) \\
m_b(A_b-\mu\tan\beta) & (\sq=\sb) \end{array} \right. \, .
\eea
As usually, we
introduce a gluon mass $\lambda$ for the regularization of the
infrared divergence. Here we define the functions
$B_0$, $B_1$, $C_0$, $C_1$, and $C_2$
as in \cite{10,11}:
\bea
\left[B_0, k^\mu B_1\right](k^2, m_0^2, m_1^2)&=&
\int\frac{d^Dq}{i\pi^2}\frac{\left[1, q^\mu\right]}
{(q^2-m_0^2)((q+k)^2-m_1^2)}\\
\left[C_0, k_t^{\mu}C_{1}-k_\bbar^{\mu}C_{2}\right](m_0^2, m_1^2, m_2^2) &=&
\int\frac{d^Dq}{i\pi^2}\frac{\left[1, q^{\mu}\right]}
{(q^2-m_0^2)((q+k_t)^2-m_1^2)
((q-k_\bbar)^2-m_2^2)}\, .\nonumber
\eea
\mn Here $k_t$ and $k_\bbar$ are the external momenta of $t$ and $\bbar$,
respectively. The parameter $r$ in eq.~(\ref{vertexcorr}) and following
equations shows the dependence on the regularization: $r = 1$ for dimensional
regularization and $r = 0$ for the dimensional reduction (DR) \cite{12}.
The dependence on $r$, however, disappears in our final result.\\

Now we turn to the quark wave--function renormalization due to the
graphs of
Fig.~1c. The two--point vertex function for $\bar{q} q$ can be written as:
\beq
\ksla(1+\Pi^q_L(k^2)P_L+\Pi^q_R(k^2)P_R)-(m_q+\Sigma^q_L(k^2)P_L
+\Sigma^q_R(k^2)P_R).
\eeq
Here we have $\Sigma^q_L(k^2)=\Sigma^q_R(k^2)\equiv\Sigma^q(k^2)$.
The correction to the amplitude from the wave--function
renormalization has the form:
\beq \label{wavecorr}
\begin{array}{c}
\delta (Y_1 P_R + Y_2 P_L)^{(w)} =
-\frac{1}{2}(\Pi^t_L(m_t^2)+\Pi^b_R(m_b^2))y_1P_R
-\frac{1}{2}(\Pi^t_R(m_t^2)+\Pi^b_L(m_b^2))y_2P_L\\
+ (m_t\dot{\Sigma}^t(m_t^2)-m_t^2\dot{\Pi}^t(m_t^2)
+ m_b\dot{\Sigma}^b(m_b^2)-m_b^2\dot{\Pi}^b(m_b^2))(y_1P_R+y_2P_L)\, ,
\end{array}
\eeq
with $\dot{\Pi}^q \equiv \frac 1 2 (\dot{\Pi}^q_L + \dot{\Pi}^q_R)$ and
$\dot X \equiv \frac{d X}{d k^2}$.
The explicit calculation yields:
\bea
(\Pi^q_L(k^2)P_L+\Pi^q_R(k^2)P_R)^{(g)}&=&
\frac{\alpha_sC_F}{4\pi}[-2B_1(k^2,m_q^2,\lambda^2) -r]\, ,\\
(\Pi^q_L(k^2)P_L+\Pi^q_R(k^2)P_R)^{(\sg)}&=&
-\frac{\alpha_sC_F}{4\pi}\left[
 2(\cos^2\theta_\sq P_L+\sin^2\theta_\sq P_R)B_1(k^2,m_{\sg}^2,m_{\sq_1}^2)
\right.\nonumber\\ &&
\mbox{\hphantom{12345}}
+2(\sin^2\theta_\sq P_L+\cos^2\theta_\sq P_R)B_1(k^2,m_{\sg}^2,m_{\sq_2}^2)
\left. \right]\nonumber\, ,
\eea
\bea
\Sigma^q(k^2)^{(g)}&=&
\frac{\alpha_sC_F}{4\pi}m_q [4 B_0(k^2,m_q^2,\lambda^2) -2 r]\, ,\\
\Sigma^q(k^2)^{(\sg)}&=&
\frac{\alpha_sC_F}{4\pi}
[m_{\sg}\sin 2\theta_\sq
(B_0(k^2,m_{\sg}^2,m_{\sq_1}^2)-B_0(k^2,m_{\sg}^2,m_{\sq_2}^2))]\,.
\eea
Finally, there are additional corrections $\delta Y_i^{(0)}$ by the
renormalization of the
quark masses in the couplings of eq.~(\ref{treecouplings}) (In the
$\overline{\mbox{DR}}$ scheme equivalent corrections are necessary as
one takes the physical masses of the quarks as input):
\bea
\delta Y_1^{(0)} = \delta y_1 &=&
\frac{g}{\sqrt{2}m_W} \delta m_b \tan\beta\nonumber\, ,\\
\delta Y_2^{(0)} = \delta y_2 &=&
\frac{g}{\sqrt{2}m_W} \delta m_t \cot\beta \label{0corr}\, ,
\eea
\bea
\mbox{with}\qquad
\delta m_q &=& \delta m_q^{(g)} + \delta m_q^{(\sg)} \, ,\nonumber\\
\delta m_q^{(g)} &=& -\frac{\alpha_sC_F}{4\pi}
[2m_q(B_0(m_q^2, 0, m_q^2)-B_1(m_q^2, 0, m_q^2) -\frac r 2)]
\nonumber\, ,\mbox{ and}\\
\delta m_q^{(\sg)} &=& -\frac{\alpha_sC_F}{4\pi}
\left[\right.\sin 2\theta_\sq m_{\sg}(B_0(m_q^2, m_{\sg}^2, m_{\sq_1}^2)
-B_0(m_q^2, m_{\sg}^2, m_{\sq_2}^2))\nonumber\\
&&\mbox{\hphantom{12345789100}}+m_q(B_1(m_q^2, m_{\sg}^2, m_{\sq_1}^2)
+B_1(m_q^2, m_{\sg}^2, m_{\sq_2}^2))\left.\right] \, .
\eea
\mn Taking all contributions eqs.~(\ref{treecouplings}, \ref{vertexcorr},
\ref{wavecorr}, \ref{0corr})
together, we get the one--loop
corrected couplings $Y_i = y_i + \delta Y_i =
y_i + \delta Y_i^{(0)} + \delta Y_i^{(v)} +
\delta Y_i^{(w)}$ with contributions
to $\delta Y_i^{(0), (v), (w)}$ from gluon and gluino exchanges
(see eq.~(\ref{2})). It can
be readily seen that they are ultraviolet finite but still infrared divergent.
The one--loop corrected decay width to ${\cal O}(\alpha_s)$ is then given
by
\bea
\Gamma(H^+\to t\bbar) &=& \frac{N_C\kappa}{16\pi m_{H^+}^3}\left[
(m_{H^+}^2-m_t^2-m_b^2)\left( y_1^2 + y_2^2 +
2 y_1{\rm Re}(\delta Y_1) +  2 y_2{\rm Re}(\delta Y_2)\right)\right.\nonumber\\
&& \left. - 4 m_t m_b \left( y_1 y_2 +
y_1{\rm Re}(\delta Y_2) +  y_2{\rm Re}(\delta Y_1)\right)\right] \, .
\eea

\section{Inclusion of the gluon emission}
For the cancellation of the infrared divergencies ($\lambda \to 0$) it is
necessary to include the ${\cal O}(\alpha_s)$ contribution from real gluon
emission as shown in Fig.~1d.\\
\noindent The decay width of $H^+\to t+\bbar+g$ is given by
\bea
&&\Gamma(H^+\to t\bbar g) = \frac{\alpha_sC_FN_C}{4\pi^2 m_{H^+}}
\left[ (y_1^2+y_2^2)\{J_1 -(m_{H^+}^2-m_t^2-m_b^2)J_2\right.
\\
&&\mbox{\phantom{1}}+ (m_{H^+}^2-m_t^2-m_b^2)^2 I_{12}\}
+4 m_t m_b y_1 y_2\{J_2 - (m_{H^+}^2-m_t^2-m_b^2)I_{12}
\left.\} \right]\, , \nonumber
\eea
with the integrals
\bea
I_{12} &=& \frac{1}{4 m_{H^+}^2}\left[ -2 \ln\left(\frac{\lambda m_{H^+}
m_t m_b}{\kappa^2}\right) \ln\beta_0 + 2 \ln^2\beta_0 - \ln^2\beta_1
-\ln^2\beta_2\right.\nonumber\\
&& \left. + 2\mbox{Sp}(1-\beta_0^2) - \mbox{Sp}(1-\beta_1^2)
- \mbox{Sp}(1-\beta_2^2)\right]\\
J_1 &=& \frac{1}{2}I^2_1+\frac{1}{2}I^1_2+I =-\frac{1}{2}I^0_1
-\frac{1}{2}I^0_2\nonumber\\
&=& \frac{1}{8 m_{H^+}^2}\left[(\kappa^2+ 6 m_t^2 m_b^2) \ln\beta_0 -\frac 3 2
\kappa (m_{H^+}^2 - m_t^2 - m_b^2)\right]\\
J_2 &=& m_t^2I_{11}+m_b^2I_{22}+I_1+I_2 \nonumber\\
&=& -\frac{1}{4 m_{H^+}^2}\left[2 \kappa \ln\left(\frac{\lambda m_{H^+}
m_t m_b}{\kappa^2}\right)+ 4 \kappa +(m_{H^+}^2 + m_t^2+m_b^2)\ln\beta_0
\right.\nonumber\\&&\left.
+2 m_t^2\ln\beta_1 +  2 m_b^2\ln\beta_2\right]\, .
\eea
Here
\bea
\beta_0\equiv\frac{m_{H^+}^2-m_t^2-m_b^2+\kappa}{2m_tm_b} &,&
\beta_1\equiv\frac{m_{H^+}^2-m_t^2+m_b^2-\kappa}{2m_{H^+}m_b} \,, \nonumber\\
\beta_2\equiv\frac{m_{H^+}^2+m_t^2-m_b^2-\kappa}{2m_{H^+}m_t} &,&
{\rm Sp}(x)=-\int_0^x\frac{dt}{t}\ln(1-t)\, ,
\eea
and $\kappa = \kappa(m_{H^+}^2, m_t^2, m_b^2)$.
The definitions and the explicit forms of
the $I$'s are given in \cite{10}.\\
The one--loop corrected decay width to ${\cal O}(\alpha_s)$ including the
real gluon emission can be written as:
\bea \label{27}
\Gamma^{\mbox{corr}}(H^+\to t\bbar+t\bbar g) &\equiv &
\Gamma(H^+\to t\bbar) + \Gamma(H^+\to t\bbar g)\nonumber\\  & = &
\Gamma^{\mbox{tree}}(H^+\to t\bbar)
+ \delta \Gamma (\mbox{gluon}) + \delta \Gamma (\mbox{gluino})\, ,
\eea
with $\Gamma^{\mbox{tree}}$given by eq.~(\ref{gammatree}), and
\bea
\delta\Gamma(\mbox{gluon})
& = &\frac{N_C\kappa}{16\pi m_{H^+}^3}\left[2 (m_{H^+}^2-m_t^2-m_b^2)
\left(y_1{\rm Re}(\delta Y_1^{(g)}) +  y_2{\rm Re}(\delta Y_2^{(g)})\right)
\right.\\&&\mbox{\hphantom{12345}}-4m_t m_b \left.\left(
y_1{\rm Re}(\delta Y_2^{(g)}) +  y_2{\rm Re}(\delta Y_1^{(g)})\right)\right]
 + \Gamma(H^+\to t\bbar g)\, ,\nonumber\\
\delta\Gamma(\mbox{gluino})
& = &\frac{N_C\kappa}{16\pi m_{H^+}^3}\left[2 (m_{H^+}^2-m_t^2-m_b^2)
\left(y_1{\rm Re}(\delta Y_1^{(\sg)}) +  y_2{\rm Re}(\delta Y_2^{(\sg)})\right)
\right.\nonumber\\&&\mbox{\hphantom{12345}}-4m_t m_b \left.\left(
y_1{\rm Re}(\delta Y_2^{(\sg)})+ y_2{\rm Re}(\delta Y_1^{(\sg)})\right)\right]
\, .
\eea
We have checked that the corrected width of eq.~(\ref{27}) is infrared finite.

\section{Numerical Results and Discussion}

We now turn to the numerical evaluation of the corrected width eq.(\ref{27}).
As the standard QCD corrections
have already been calculated \cite{7}, it is
interesting here
to study the influence of the gluino (and $\tilde{t}_i , \tilde{b}_j$) exchange
corrections $\delta\Gamma$(gluino). The whole analysis depends on the
following parameters defined at the weak scale:
$m_{H^+}, \tan\beta , \mu , A_t , A_b$, $M_{\tilde{Q}},
M_{\tilde{U}}, M_{\tilde{D}}$, and $m_{\tilde{g}}$. For simplicity we assume
$M_{\tilde{Q}} = M_{\tilde{U}} = M_{\tilde{D}}$ and $A_t = A_b \equiv A$.
We have found that our final results are rather insensitive to these
assumptions. We take $m_t = 180$~GeV, $m_b = 5$~GeV, $m_W = 80$~GeV,
$m_Z = 91.2$~GeV, $\sin^2 \theta_W = 0.23$, $g^2/(4\pi) = \alpha_2 =
\alpha/\sin^2 \theta_W = 0.0337$ and $\alpha_s = \alpha_s(m_{H^+})$. We use
$\alpha_s(Q) = 12 \pi /\{(33 - 2 n_f)\ln(Q^2/\Lambda^2_{n_f})\}$ with
$\alpha_s(m_{Z}) = 0.12$ and the number of quark flavors $n_f = 5 (6)$ for
$m_b < Q \leq m_t$ (for $Q > m_t$).\\

In Fig. 2 we show the dependence of $\delta\Gamma$(gluino) as a
function of $A$ and $M_{\tilde{Q}}$ for $\tan\beta = 2$ (a) and $12$ (b), and
$(m_{H^+}, m_{\sg}, \mu) = (400, 550, 300)$~(GeV). We see that the
size of the
SUSY--QCD correction $\delta\Gamma(\mbox{gluino})$ can be large going up to
$\sim$ 50\% and that it can be comparable to or even larger than the
standard QCD correction $\delta\Gamma(\mbox{gluon})$ in a large parameter
region.
For fixed
$\tan\beta$, $\delta\Gamma(\mbox{gluino})$
has a strong dependence on the parameters $M_{\tilde{Q}}$ and
$A$ which determine the masses and couplings of $\tilde{t}_{1,2}$ and
$\tilde{b}_{1,2}$. $\delta\Gamma$(gluino) is smaller for larger masses of
$\tilde{t}_1$ and $\tilde{b}_1$:
for $\tan\beta=2$, the correction due to $\delta\Gamma({\rm gluino})$ is
about $-15$\% for
$(M_{\tilde{Q}},A)=$(100 GeV, 300 GeV) (where $m_{\tilde{t}_1}
\simeq 119$~GeV, and $m_{\tilde{b}_1}\simeq 98$GeV), but it is still
$\sim -5$\% for
larger squark masses $(M_{\tilde{Q}},A)=$(400 GeV, 300 GeV)
(where $m_{\tilde{t}_1}\simeq 405$ GeV, and $m_{\tilde{b}_1}\simeq 399$GeV).
This tendency is consistent with the decoupling theorem for the MSSM.
Notice also the
different behaviour for $\tan\beta = 2$ and $\tan\beta = 12$.\\

In Fig.~3 we show the $m_{H^+}$ dependence of
$\Gamma^{\rm{tree}}$, $\Gamma^{\rm{tree}} + \delta\Gamma$(gluon), and
$\Gamma^{\rm{corr}} = \Gamma^{\rm{tree}} + \delta\Gamma$(gluon) +
$\delta\Gamma$(gluino) for
$\tan\beta = 2$ (a) and $12$ (b), and
$(m_{\sg}, \mu, M_{\tilde Q}, A) = (400, -300, 200, 200)$~(GeV).
The parameter values
correspond to fixed stop and sbottom masses:
$m_{\tilde{t}_1} = 90$~GeV, $m_{\tilde{t}_2} = 366$~GeV, $m_{\tilde{b}_1}
= 193$~GeV, and $m_{\tilde{b}_2} = 213$~GeV (for $\tan\beta = 2$) and
$m_{\tilde{t}_1} = 173$~GeV, $m_{\tilde{t}_2} = 333$~GeV,
$m_{\tilde{b}_1}
= 152$~GeV, and $m_{\tilde{b}_2} = 247$~GeV (for $\tan\beta = 12$).
(Note that for $m_{\sg} = 400$~GeV the D0 mass limit of the mass--degenerate
squarks of five flavors (excluding $\st_{1,2}$) is $m_{\sq} \gsim 140$~GeV
\cite{13}.)
We see again that the correction
$\delta\Gamma$(gluino) can be quite large and that it is comparable to or
even larger than
$\delta\Gamma$(gluon)
in a large region. Quite generally, the corrections $\delta\Gamma$(gluon)
and $\delta\Gamma$(gluino) are bigger for larger $\tan\beta$, but
it can happen that they
partly cancel each other. The correction $\delta\Gamma$(gluon) has
already been calculated in \cite{7}. Our results on
$\delta\Gamma$(gluon) agree numerically
with ref.~7 within 10\%.\\

In Fig.~4 we show a contour--plot for $\frac{\delta\Gamma
\rm{(gluino)}}{\Gamma^{\rm{corr}}}$ in the $\tan\beta - m_{\tilde{g}}$
plane
for $(m_{H^+},\mu, M_{\tilde Q}, A) = (400,-300, 250, 300)$~GeV.
This correction rises with increasing $\tan\beta$, going up to 50\%!
Concerning the $m_{\tilde{g}}$
dependence, $\frac{\delta\Gamma
\rm{(gluino)}}{\Gamma^{\rm{corr}}}$ increases up to $m_{\tilde{g}} = 300 -
450$~GeV and then decreases gradually as $m_{\tilde{g}}$ increases.
It is striking that even for a large gluino
mass ($\sim$ 1 TeV) $\frac{\delta\Gamma\rm{(gluino)}}{\Gamma^{\rm{corr}}}$ is
larger than 10\% for $\tan\beta \gsim$~3. From  Figs.~2 and 4 we see
that the
correction $\delta\Gamma$(gluino) decreases much faster for increasing
$M_{\tilde Q}$ than for increasing $m_\sg$.\\

In Fig.~5 we show contour lines of $\delta\Gamma\rm{(gluino)}$ in the
$\mu - A$ plane for $\tan\beta = 2$~(a) and 12 (b), and $(m_{H^+},
m_{\tilde{g}}, M_{\tilde Q}) = (400, 550, 300)$~GeV. This correction has a
strong dependence on $\mu$ and a significant dependence on $A$. We have
found that the $A$~\mbox{dependence}
for $\tan\beta = 1$ is much stronger than that
for $\tan\beta = 2$.\\

The reason for the large contribution of $\delta\Gamma\rm{(gluino)}$ as
compared to $\delta\Gamma\rm{(gluon)}$ is the following: The vertex--correction
part of the gluino--exchange [gluon--exchange] corrections (see Fig.~1b and
eq.~(5)) is proportional to the $H^+ \bar{\st} \sb$ coupling [$H^+ \bar{t} b$
coupling] which is essentially $\sim (A_t + \mu \tan\beta) h_t \cos\beta
+ (A_b + \mu \cot\beta) h_b \sin\beta$ [$\sim h_t \cos\beta + h_b \sin\beta$].
Hence the vertex--correction part of the gluino--exchange corrections
$\delta\Gamma\rm{(gluino)}$ can be strongly enhanced relative to that of the
gluon--exchange corrections $\delta\Gamma\rm{(gluon)}$ in the case the
$\tilde{q}$--mixing parameters $A$ and $\mu$ are large. In this case $\st_1$
and $\sb_1$ tend to be light due to a large mass--splitting. Note that the
$\sb$--mixing effect plays a very important role for large $\tan\beta$.

\section{Conclusion}

Summarizing, we have performed a complete calculation of the
${\cal O}(\alpha_s)$ QCD corrections to the width of $H^+ \to t \bar{b}$ within
the MSSM. We have found that the QCD corrections are quite important. A
detailed
numerical analysis has shown
that the SUSY--QCD corrections (due to gluino, $\st$ and
$\sb$ exchanges) can be comparable to or even larger than the standard QCD
corrections in a large region of the MSSM parameter space; here the mixings
of $\st_L - \st_R$ and $\sb_L - \sb_R$ play a crucial role. This could
significantly affect the phenomenology of the $H^+$ search.\\
After having finished this study, we have been informed on a recent paper
\cite{14} dealing with the same subject.

\section*{Acknowledgements}

The work of Y.Y. was supported in part by the Fellowships of the Japan
Society for the Promotion of
Science and the Grant-in-Aid for Scientific
Research from the Ministry of Education, Science and Culture of Japan, No.
06-1923 and 07-1923.
The work of A.B., H.E., and W.M. was supported by the ``Fonds zur
F\"orderung der
wissenschaftlichen Forschung'' of Austria, project no. P10843-PHY.


\vspace{20mm}
\section*{Figure Captions}
\renewcommand{\labelenumi}{Fig.~\arabic{enumi}} \begin{enumerate}

\vspace{6mm}
\item
All diagrams relevant for the calculation of the
${\cal O}(\alpha_s)$ QCD corrections to the width of $H^+ \to t \bar{b}$ in the
MSSM.

\vspace{6mm}
\item
Contour lines of $\delta\Gamma\rm{(gluino)}$~(GeV) in the $A-M_{\tilde Q}$
plane
for $\tan\beta = 2$ (a) and $12$ (b), and
$(m_{H^+}, m_{\sg}, \mu) = (400, 550, 300)$~(GeV).
For these parameter values one
has ($\Gamma^{\rm{tree}}$~(GeV), $\delta\Gamma(\rm{gluon})$~(GeV)) = $(4.10,
0.31)$ and $(1.91, -0.66)$ for Figs.~2a and 2b, respectively.
The shaded area is excluded by the LEP bounds $m_{\st_1,\sb_1} \gsim 45$~GeV.
Note that for $m_{\sg} \simeq 550$~GeV one has no squark mass bound from D0
experiment \cite{13}.

\vspace{6mm}
\item
$m_{H^+}$ dependence of $\Gamma^{\rm{tree}}$ (dashed line),
$\Gamma^{\rm{tree}} + \delta\Gamma$(gluon) (dot--dashed line), and
$\Gamma^{\rm{corr}} = \Gamma^{\rm{tree}} + \delta\Gamma$(gluon) +
$\delta\Gamma$(gluino) (solid line) for $\tan\beta = 2$ (a) and $12$ (b), and
$(m_{\sg}, \mu, M_{\tilde Q}, A) = (400, -300, 200,200)$~(GeV).

\vspace{6mm}
\item
Contour lines of $\delta\Gamma$(gluino)/$\Gamma^{\rm{corr}}$ in the
$\tan\beta$--$m_{\tilde{g}}$ plane for
$(m_{H^+}, \mu, M_{\tilde Q}, A) = (400, -300, 250, 300)$~(GeV).
The area below the dotted line is excluded by the LEP limit
$m_{{\tilde\chi}^+_1}
\gsim 45$~GeV (assuming $m_{\sg} = (\alpha_s/\alpha_2) M_2 \simeq 3.56 M_2$),
where $\alpha_2 = g^2/(4\pi)$, $M_2$ is the SU(2) gaugino mass, and
$m_{{\tilde{\chi}}^+_1}$ is the lighter chargino mass.

\vspace{6mm}
\item
Contour lines of $\delta\Gamma$(gluino)~(GeV) in the $\mu - A$ plane for
$\tan\beta = 2$ (a) and 12 (b), and $(m_{H^+},
m_{\tilde{g}}, M_{\tilde Q}) = (400, 550, 300)$~GeV.
For these parameter values one has
($\Gamma^{\rm{tree}}$~(GeV), $\delta\Gamma(\rm{gluon})$~(GeV)) = $(4.10,
0.31)$ and $(1.91, -0.66)$ for Figs.~5a and 5b, respectively. The shaded
area is excluded by the LEP limits $m_{\st_1, \sb_1, {\tilde{\chi}}_1^+} \gsim
45$~GeV. For $m_{\sg} \simeq 550$~GeV one has no $m_{\sq}$ limit from the D0
experiment \cite{13}.

\end{enumerate}

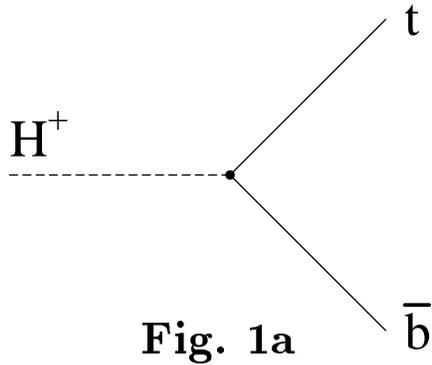
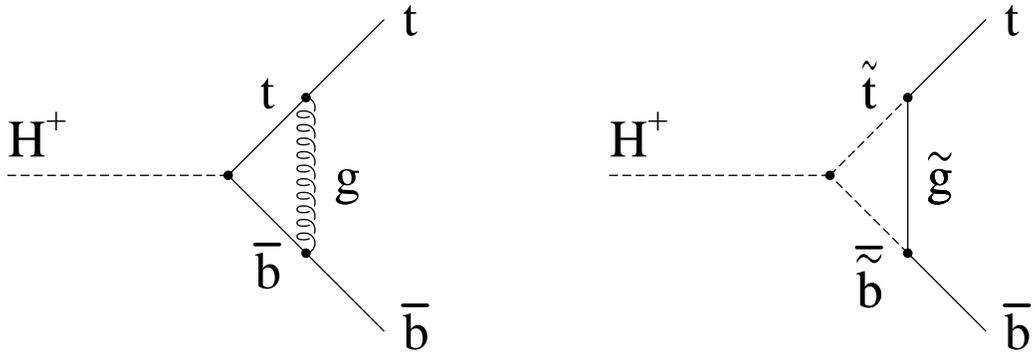
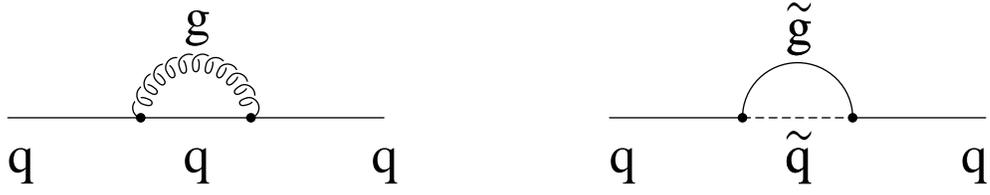
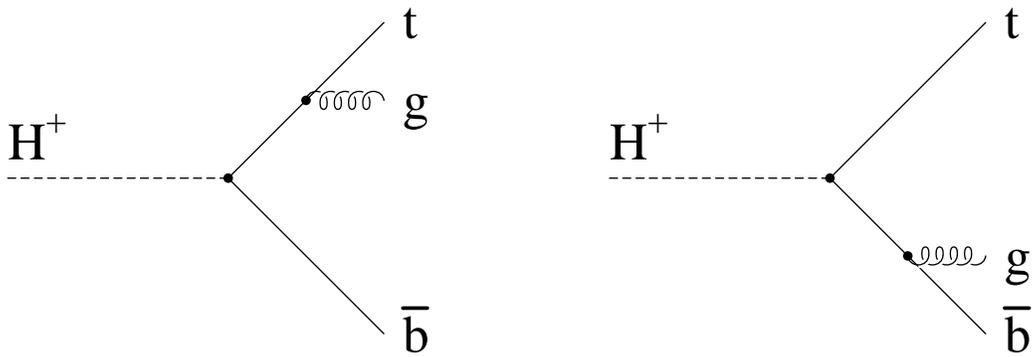
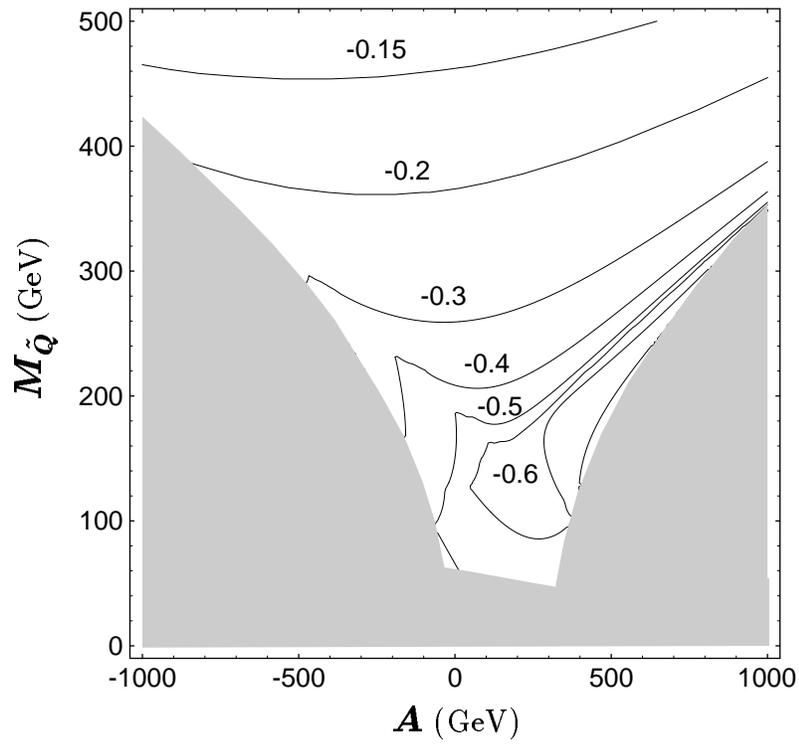
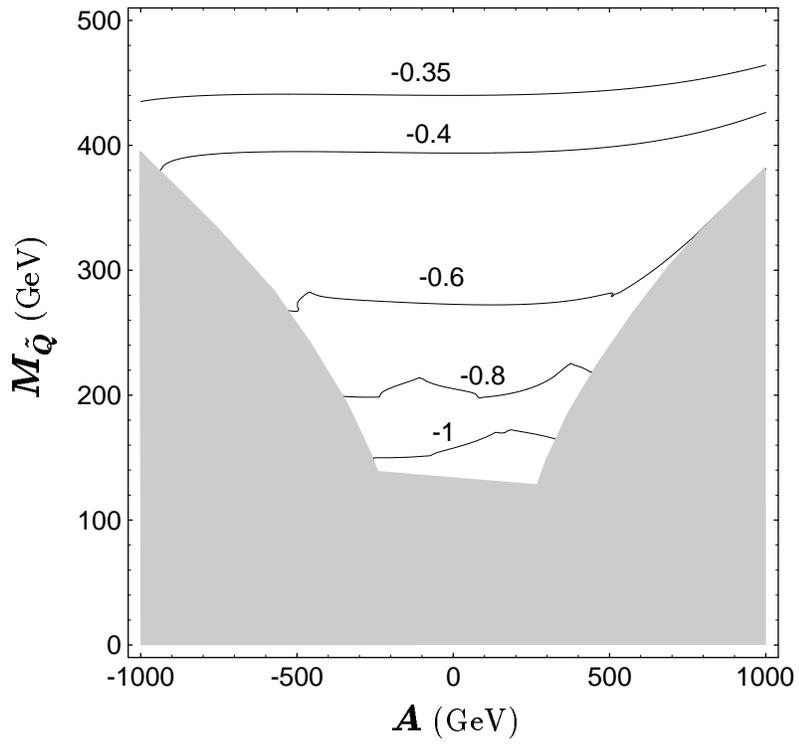
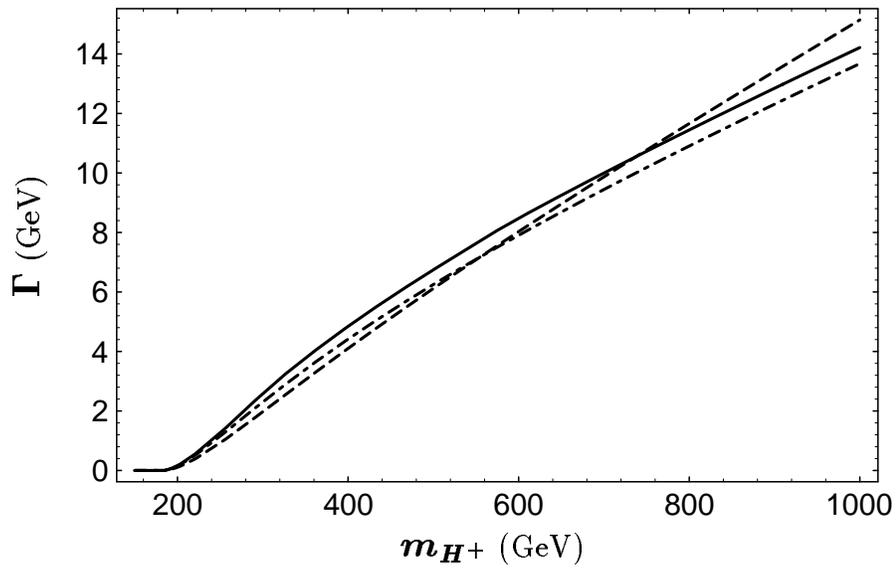
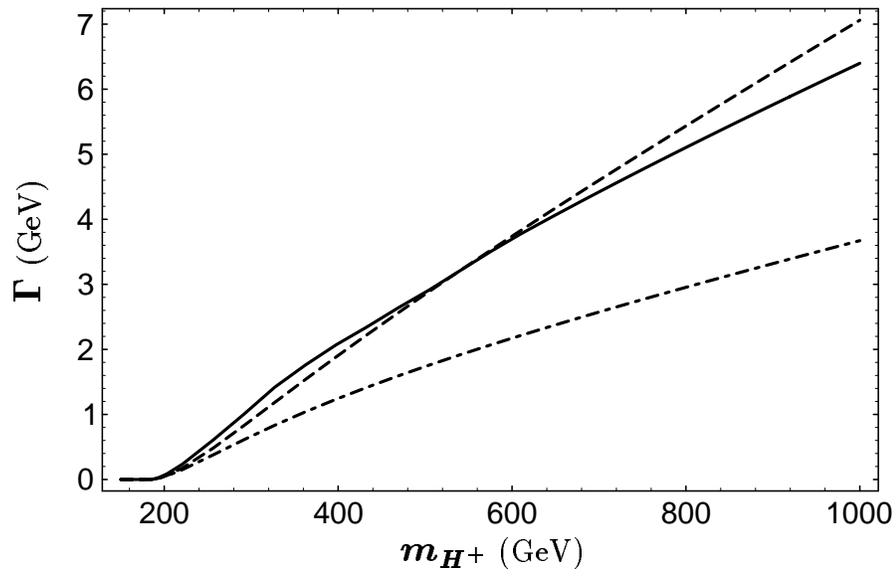
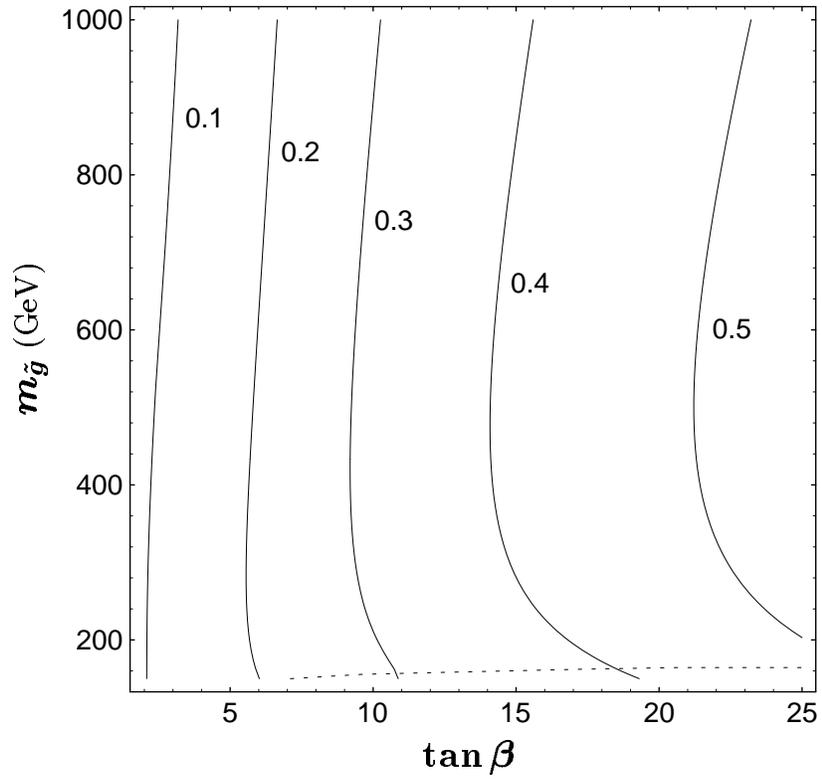
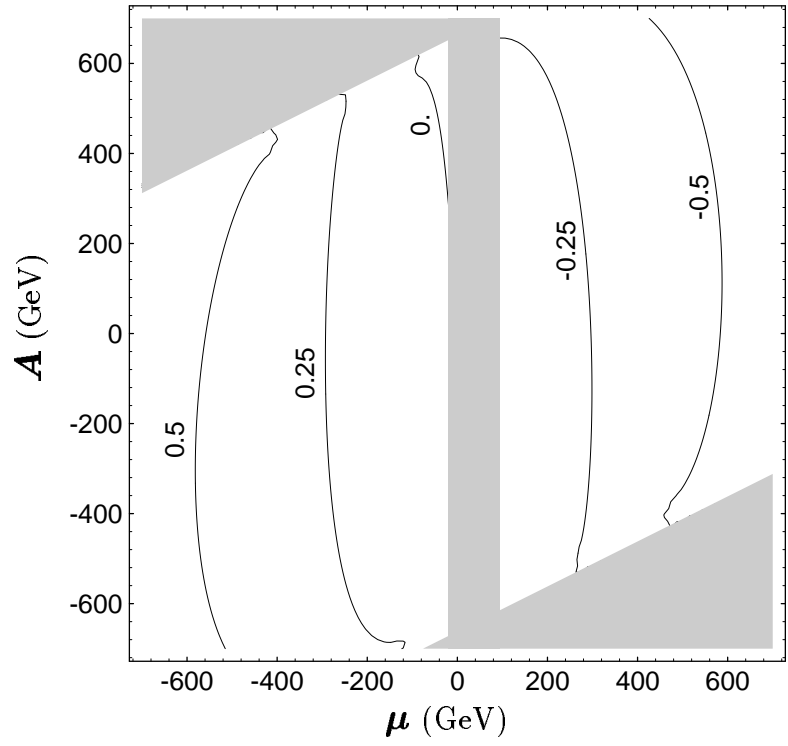
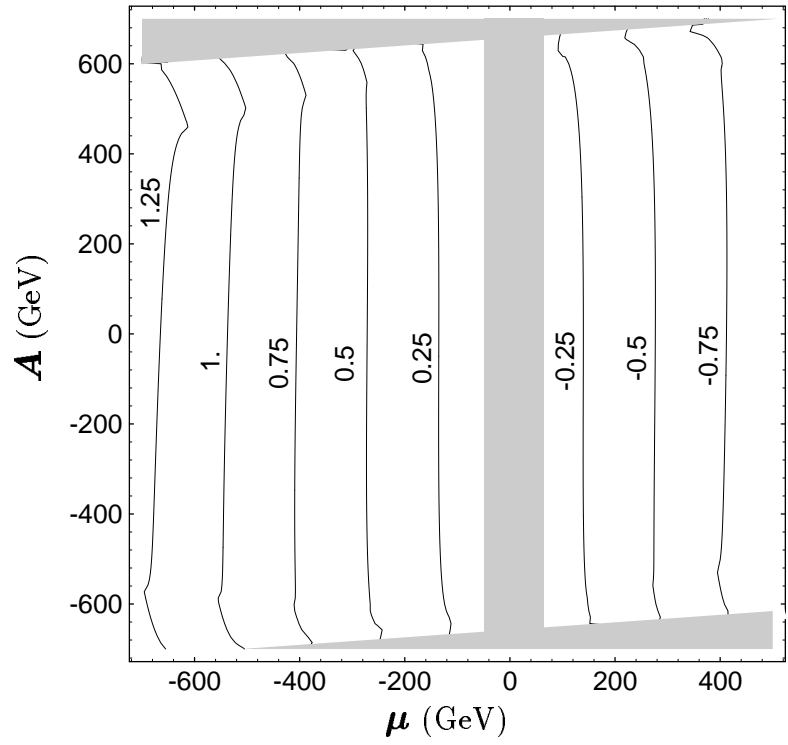
\end{document}